\documentclass[sn-mathphys,Numbered]{sn-jnl}

\usepackage{graphicx}%
\usepackage{multirow}%
\usepackage{amsmath,amssymb,amsfonts}%
\usepackage{amsthm}%
\usepackage{mathrsfs}%
\usepackage[title]{appendix}%
\usepackage{xcolor}%
\usepackage{textcomp}%
\usepackage{manyfoot}%
\usepackage{booktabs}%
\usepackage{algorithm}%
\usepackage{algorithmicx}%
\usepackage{algpseudocode}%
\usepackage{listings}%

\theoremstyle{thmstyleone}%
\theoremstyle{thmstyletwo}%

\theoremstyle{thmstylethree}%

\begin{document}



\title{Observation of topologically protected compact edge states in flux-dressed graphene photonic lattices}

\author[1]{Gabriel C\'aceres-Aravena}
\email{gabriel.caceres-aravena@uni-rostock.de}
\equalcont{These authors contributed equally to this work.}

\author[2]{Milica Nedi\'c}
\email{milica.brankovic@vin.bg.ac.rs}
\equalcont{These authors contributed equally to this work.}

\author[3]{Paloma Vildoso}
\email{palomavildoso1@gmail.com}
\equalcont{These authors contributed equally to this work.}

\author[2]{Goran Gligori\'c}
\email{goran79@vin.bg.ac.rs}

\author[2]{Jovana Petrovic}
\email{jovanap@vin.bg.ac.rs}

\author[3]{Rodrigo A. Vicencio}
\email{rvicencio@uchile.cl}

\author*[2]{Aleksandra Maluckov}
\email{sandram@vin.bg.ac.rs}

\affil[1]{\orgdiv{Institute of Physics}, \orgname{University of Rostock}, \city{Rostock}, \postcode{18051}, \country{Germany}}

\affil[2]{\orgdiv{Vin\v ca Institute of Nuclear Sciences, National Institute of the Republic of Serbia}, \orgname{University of Belgrade}, \orgaddress{\street{P.O.B. 522}, \city{Belgrade}, \postcode{11001},  \country{Serbia}}}

\affil[3]{\orgdiv{Departamento de Física and Millenium Institute for Research in Optics–MIRO}, \orgname{Facultad de Ciencias Físicas y Matemáticas, Universidad de Chile}, \orgaddress{\city{Santiago}, \postcode{8370448},  \country{Chile}}}

\date{\today}


\maketitle

\begin{center}
\textbf{Abstract}
\end{center}
\begin{abstract}

Systems with engineered flatband spectra are a postulate of high-capacity transmission links and a candidate for high-temperature superconductivity. However, their operation relies on the edge or surface modes susceptible to fluctuations and fabrication errors. While the mode robustness can be enhanced by a combination of Aharonov-Bohm caging and topological insulation, the design of the corresponding flatbands requires approaches beyond the standard $k$-vector-based methods. Here, we propose a synthetic-flux probe as a solution to this problem and a route to the realization of ultra-stable modes. We prove the concept in a laser-fabricated graphene-like ribbon photonic lattice with the band-flattening flux induced by ``P'' waveguide coupling. The topological non-triviality is witnessed by an integer Zak phase derived from the mean chiral displacement. Mode stability is evidenced by excellent mode localization and the robustness to fabrication tolerances and variations of the input phase. Our results can serve as a basis for the development of multi-flat-band materials for low-energy electronics.

\end{abstract}

\section{INTRODUCTION}\label{sec:Results}
\vskip 0.25cm
The quest for materials with superconducting properties at convenient operating temperatures has gained new momentum with the possibility to investigate such materials in affordable optics experiments, notably in photonic lattices~\cite{tcs, photcr, tph, pti, roadmap}. Departing from the traditional Bardeen–Cooper–Schrieffer (BCS) theory, it has been hypothesized that the electron pairing around dispersionless energy bands results in superconductivity with the critical temperature linearly proportional to the electron interaction strength\cite{volovik2015}. However, dispersionless bands, commonly known as flat bands (FBs), store energy in the modes confined to edges, surfaces or interfaces. This makes them highly sensitive to environmental influences and fabrication errors~\cite{bcs1,PhysRevB93201116,bcs2}. Stability can be enhanced by topological insulation achieved by an interplay between the Aharonov-Bohm caging and topological protection, such as that realized in Creutz-like ladders~\cite{zero,zero1,Topology_1d_SP}. 
However, $k$-independence of FBs does not permit determination of the topological invariants, neither directly, based on the Berry-curvature, nor indirectly, based on analogy with the Su-Schrieffer–Heeger (SSH) model \cite{multissh,azboth}. Exceptions are the configurations which allow Hamiltonian reduction to the SSH Hamiltonians \cite{zero}. All this makes the design of ultra-robust edge states with the FB spectrum difficult. 

In response to these challenges, we design a graphene-like photonic ribbon whose topological properties can be controlled by synthetic flux. The all FB ribbon spectrum is achieved for $\pi$-flux, which is realized by inserting only S or S and P waveguides in a particular order through the lattice, \cite{grapheneSP}. 
The induced transitions between trivial and nontrivial topological phases enable probing of the dispersionless band topology and determination of the topological invariants. Topological transitions are detectable by standard means: two-band SSH-based graphical method, mean chiral displacement (MCD) \cite{chiral1,chiral2,chiral3} and numerical projector method \cite{book,azboth,leaky}. While a range of fluxes drives the system into the topologically nontrivial phase, a particular flux value, $\Phi=\pi$, yields destructive interference and ultimate band-flattening. The corresponding edge modes are ultra-protected by the synergy of Aharonov-Bohm (AB) and topological effects.  

For verification of the proposed flux method, we use the femtosecond (fs) laser writing technique~\cite{alexfs} to inscribe a lattice in the form of a quasi-1D graphene ribbon in armchair configuration~\cite{bauer-1,bauer-2}. By determining the MCD from the experimental data, we confirm the topological nontriviallity of the central bands and the corresponding edge modes. We demonstrate a zero-mode edge state and its robustness to the input-state phase variations and fabrication tolerances. To the best of our knowledge, this is the first complete experimental evidence of an ultra-robust and perfectly compact topologically nontrivial edge state in a one-dimensional lattice.  

\begin{figure}
\includegraphics[width=1.0\textwidth]{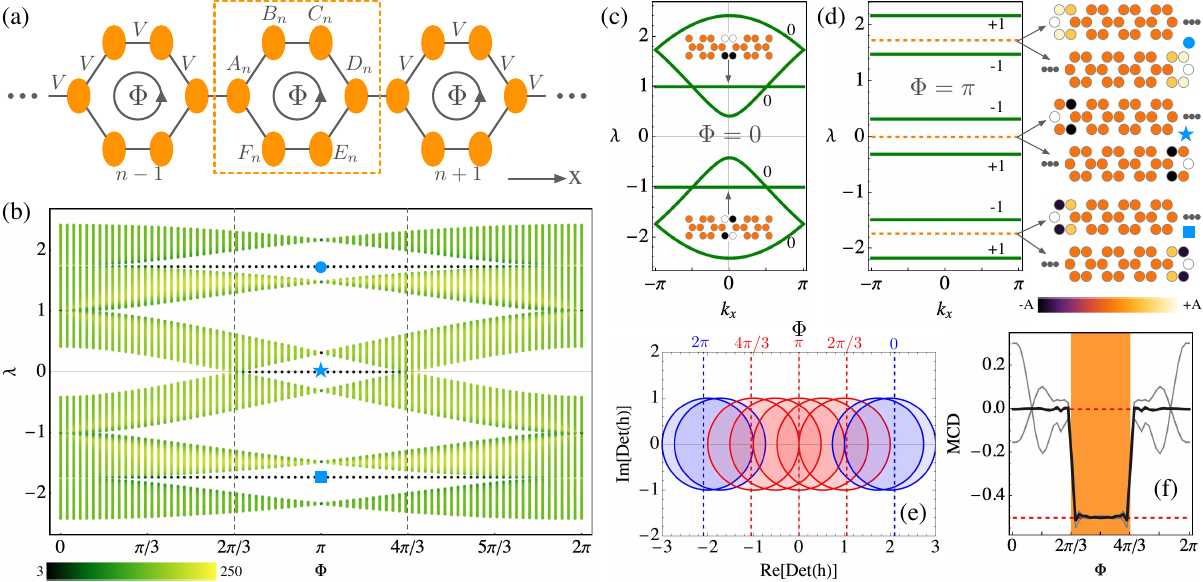}
   \caption{(a) Schematic graphene-like ribbon lattice including a flux $\Phi$ on every hexagonal ring. (b) Energy-band spectrum of a finite lattice ($N=50$ unit cells) as a function of flux $\Phi$. The topologically non-trivial domain $\Phi\in[2\pi/3,4\pi/3]$ is bounded by vertical dashed lines. (c) and (d) show the band spectra $\lambda$ vs. $k_x$, inside the 1st Brillouin zone (BZ), for fluxes $\Phi=0$ and $\Phi=\pi$, respectively. In (d) the edge-states correspond to horizontal dashed lines, with the respective mode profiles to the right. Energy spectra are normalized by $V=1$. (e) Determinant $\text{Det}(h)=-\cos k_x+2\cos(\Phi/2)+i\sin k_x$, of SSH-like reduced Hamiltonian $h$, in the complex plain (see Supplementary Information). The red circles wind around or touch the origin, defining the nontrivial topological cases. (f) The mean chiral displacement (MCD) as a function of $\Phi$ (black solid line). Gray lines correspond to the mean field displacements from initially excited single A, C, E sites (Supplement).}  
\label{fig:Spectrum}
\end{figure}

\section{RESULTS}\label{sec:Results}
\vskip 0.25cm
\noindent{\bf Topological invariants and lattice properties}
\vskip 0.25cm
We first present the method for probing the topological triviality of a lattice by artificial flux. We do this on a model of a hexagonal graphene-like ribbon in the armchair configuration. In the absence of flux, the underlying lattice structure is characterized by bipartite and top-bottom symmetries, in which each unit cell consists of six linearly coupled sites, as described in Figure \ref{fig:Spectrum}(a). Evolution of the complex optical field $\psi_{n}=(A_{n}\,B_{n}\,C_{n}\,D_{n}\,E_{n}\,F_{n})^T$, with $n=1,...,N$ and $N$ the number of unit cells, is governed by a set of linear coupled equations in $k_x$-space: 
\begin{equation}
i\frac{d {\tilde \psi}(k_x)}{dz}=H(k_x)\tilde{\psi}(k_x)\ ,
\label{eq:LSE}
\end{equation}
where ${\tilde \psi}(k_x)$ is the optical field in $k_x$-space, $k_x$ the transversal quasimomentum, $z$ the propagation coordinate, and $H$ the Bloch Hamiltonian or lattice coupling matrix. The presence of flux $\Phi$ breaks the symmetry and introduces imaginary components into the coupling coefficients:
\begin{equation} 
\hat{H} = V\left( \begin{array}{cccccc} 0 & e^{-i \phi} & 0 & e^{-i k_x} & 0 & e^{i \phi} \\  
e^{i \phi} & 0 & e^{-i \phi}& 0 &0 &0 \\  
0 &e^{i \phi} & 0 &e^{-i \phi}&0 &0 \\ 
e^{i k_x} & 0 &e^{i \phi}& 0 &e^{-i \phi}&0 \\
0 & 0 & 0&e^{i \phi} &0 &e^{-i \phi}
\\ 
e^{-i \phi} & 0 &0&0 & e^{i \phi}&0  \end{array} \right)\ ,
\label{eq:Hamiltonian}
\end{equation}
\noindent where $\phi=\Phi/6$. $Ve^{\pm i\phi}$ are the complex intra-cell coupling coefficients, while $Ve^{\pm ik_x}$ are the inter-cell coupling coefficients (for simplicity, we have set the unit cell distance as $1$), according to the geometry described in Fig.~\ref{fig:Spectrum}(a) (more details on equations at Supplementary material).  

We compute the eigenvalue spectrum $\{\lambda\}$ representing the longitudinal propagation constants along the $z$ direction for a finite lattice with $N=50$ unit cells, and show our results in Fig.~\ref{fig:Spectrum}(b), where darker color corresponds to a smaller participation ratio, i.e. stronger localization. 
In the absence of flux [see Fig.~\ref{fig:Spectrum}(c)], the bipartite ribbon spectrum contains an empty gap around zero, a pair of symmetric flat bands at $\lambda=\pm V$ (with the FB states as insets), and two pairs of dispersive bands~\cite{grapheneSP}. In this case, there are no isolated bands, and the topological invariant is not defined. The introduction of flux breaks the top-bottom ribbon symmetry, whereas the chiral symmetry is preserved. Singular FBs are transformed into gapped dispersive bands and an opening of the gaps between outer bands appears. The two outer gaps host a doublet of edge modes each [circle and square in Figs.~\ref{fig:Spectrum}(b) and (d)]. A further increment in flux $\Phi$ results in the closing and reopening of the central gap at $\Phi=2\pi/3$ - followed by the creation of a pair of zero ($\lambda=0$) edge modes [star in Fig.~\ref{fig:Spectrum}(b)], in analogy with the SSH system~\cite{delplace}. These degenerated edge modes decay exponentially from the respective surface and their localization increases as the central band gap increases (see Supplementary information). The spectrum symmetry dictates the reverse trend as the flux is further increased, closing the gap at $\Phi=4\pi/3$ and restoring the original trivial properties.

For a particular value of flux $\Phi=\pi$, the Aharonov-Bohm effect causes the bands to collapse into flat bands~\cite{spirodrigo}, forming a fully flat multi-band spectrum, as Figs.~\ref{fig:Spectrum}(b) and (d) clearly show. The flatness is associated with degeneration of $6(N-1)$ eigenvalues into $6$ flat bands at $\lambda=\pm2.17V,\ \pm 1.48V, \ \pm 0.31V$ with $N-1$-times degeneration each. In the gaps, in between the flat bands, the triplet of double degenerated edge modes continues to exist at $\lambda=0,\pm \sqrt{3}V$ [horizontal dashed lines and respective edge mode profiles in Fig.~\ref{fig:Spectrum}(d)]. As the result of destructive interference, all the corresponding edge mode profiles feature zero light power at the connecting lattice sites A or D. This fully prevents state transport and dispersion. Specifically, at $\Phi=\pi$, the zero edge modes are perfectly compact and formed by only three nonzero amplitude sites. These perfectly compact edge states cannot be formed by any FB states superposition~\cite{FBrep}. 

An estimate of bulk topological invariant is mandatory to ascertain the topological non-triviality of the compact edge states. However, the topological invariants elude a clear definition in the systems with FB $k$-independent spectrum. To resolve this, we tune the synthetic flux and evaluate the invariant (Zak phase in this case), in the $k$-dependent spectral region, upon which we formally converge to values at $\Phi=\pi$. Zak phase, also known as winding number (Supplementary information), provides information about the winding behavior of the energy bands in the momentum space, distinguishing between different topological phases in the lattice. Due to the bulk-edge correspondence~\cite{phothalleffect,topo_review1,topo_review2}, an integer value of Zak phase in units of $\pi$ is equal to the number of edge-mode pairs within a gap and signals topological non-triviality. Here, we evaluate the Zak phase by two methods: the geometric method developed on the SSH-analogy~\cite{azboth} and the mean chiral displacement method~\cite{chiral1,chiral2,chiral3} (results are corroborated by projector method in Supplementary information). 

Re-opening and re-closing of the central gap, accompanied by the appearance of the zero edge modes, is analogous to what is observed in the SSH model. This is corroborated by the transformation of Hamiltonian of the flux-dressed ribbon lattice into an SSH Hamiltonian parameterized by flux, as shown in Supplementary information. The winding number of the SSH Hamiltonian can be estimated by plotting its determinant in the complex plane and observing how many times it winds around the origin, see Fig.~\ref{fig:Spectrum}(e). The topological phase transition is marked by the zeros of the determinant at $\Phi=2\pi/3$ and $\Phi=4\pi/3$. The flux values within these limits render the lattice topologically non-trivial regime. For $\Phi=\pi$, the determinant draws a zero-centred circle, thus reproducing the $k_x$-invariance. 

The non-triviality of the zero edge mode is further confirmed by the MCD calculated as a displacement (in cells units) of the spatial profile from the input position~\cite{chiral1},\cite{longhiMFD},\cite{SSHdiamond}. It has been shown that in a lattice with 2 FBs, MCD value close to 1/2 signals the non-trivial topological phase~\cite{longhiMFD,SSHdiamond,longhi}. In Fig.~\ref{fig:Spectrum}(f), we plot the dependence of MCD with respect to flux $\Phi$ and observe that the MCD takes values close to -1/2 in the whole non-trivial region $\Phi\in[2\pi/3, 4\pi/3]$. This result indicates that the MCD can be used as a nontriviality parameter also in a lattice with multiple FBs. Final confirmation of the zero-edge mode topological nontriviality is achieved by applying the projector method, as described in Supplementary information.

Based on the above, we conclude that the zero edge modes are protected by both the destructive interference (asrising from Aharonov-Bohm caging) and the lattice topology. Hence, these states are perfectly localized, compact, and ultra-stable stationary solutions. Their chiral symmetry remains intact during the topological to trivial transitions.

\vskip 0.25cm
\noindent{\bf Experiments on flux-dressed graphene photonic lattices}
\vskip 0.25cm

\begin{figure}
\includegraphics[width=\textwidth]{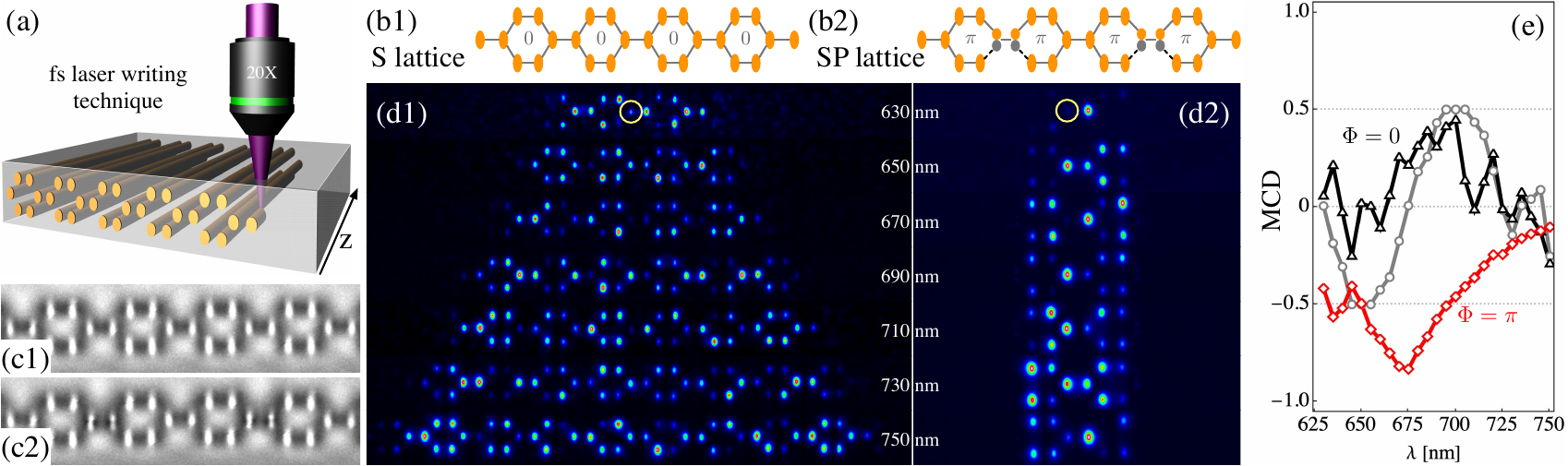}\
  \caption{Bulk excitation. (a) Illustration of the fs-laser writing technique for a graphene-like ribbon. (b1) and (b2) show S and SP configurations (fluxes denoted in each case). (c1) and (c2) show white-light microscope images for the fabricated lattices. (d1) and (d2) present output intensities after single site excitations (at sites indicated by yellow circles), on a ribbon with $\Phi=0$ and $\Phi=\pi$, respectively. The input wavelength is indicated at the center. (e) Mean chiral displacement for sites ACE versus wavelength, for trivial (black) and topological (red) lattices. Gray data in (e) shows the average MCD for CE sites only.}  
\label{fig:EdgeCompacton}
\end{figure}

We provide evidence of the trivial and topological properties of a graphene-like ribbon lattice in two experimentally available fluxes cases: $\Phi=0,\pi$. Different graphene-like photonic lattices are fabricated inside a $L=70$ mm borosilicate glass by direct laser writing~\cite{alexfs}, as sketched in Fig.~\ref{fig:EdgeCompacton}(a). If the lattice is composed of $S$ identical waveguides only, no flux ($\Phi=0$) is induced. Fig.~\ref{fig:EdgeCompacton}(b1) shows a sketch for a trivial \textit{S} lattice and Fig.~\ref{fig:EdgeCompacton}(c1) its implementation. A nontrivial flux is induced by replacing a set of $S$ waveguides with dipole-like $P$ waveguides at every other hexagon, as described in Figs.~\ref{fig:EdgeCompacton}(b2) and (c2). The presence of \textit{P} waveguides on a lattice produces the appearance of negative coupling constants and, therefore, an effective induction of flux $\Phi=\pi$ on closed rings~\cite{spirodrigo}. Effective interaction in between $S$ and $P$ modes is controlled by a tuning optimization protocol~\cite{spdimer}, which matches the propagation constants of different modes at a given wavelength. Here, the required flux $\Phi=\pi$ is optimized at a wavelength $640$ nm. 

To study the light transport through the lattices, we exploit the monotonous dependence of the coupling length on wavelength~\cite{SSHdiamond}. We capture the transport at different dynamical instants in a lattice of a fixed length, $L$, by simply varying the input wavelength (see Methods for more details). These experiments were performed using a supercontinuum (SC) laser source, in the wavelength range of $\{630,750\}$ nm, focused to excite an individual lattice site in the middle (bulk excitation) or at the edge of the lattice (edge excitation).
 
We first excite the trivial lattice at the bulk $D$ site [Fig.~\ref{fig:EdgeCompacton}(d1)] and observe a ballistic transport across the lattice, similar to 1D discrete diffraction~\cite{SSHdiamond}. In this case, the input excitation does not excite the FB modes. Therefore, only the dispersive part of the spectrum is activated. Inclusion of the FB along with the dispersive modes, e.g., by excitation of $B, C, D, E$ sites, results in a mixture of diffraction and energy oscillations strongly localized at the input ring region~\cite{grapheneSP} (graphs in Supplementary information). A strongly contrasting behaviour is observed for a nontrivial flux $\Phi=\pi$, as shown in Fig.~\ref{fig:EdgeCompacton}(d2). For an identical bulk $D$ excitation, the light cannot overcome the flux-induced effective barrier and remains oscillating on a reduced area, covering only two lattice rings. This is a result of the AB caging effect in a system with a flat-only linear spectrum [Fig.~\ref{fig:Spectrum}(d)]. 

For bulk excitation, we compute the MCD from the experimental results by averaging the excitation of A, C and E bulk lattice sites [see Fig.~\ref{fig:EdgeCompacton}(e)], consistently with the lattice chiral symmetry (see Methods). For $\Phi=0$, the MCD oscillates around zero, indicating a zero Zak phase. In this case, the light is diffracting strongly through the lattice [Fig.~\ref{fig:EdgeCompacton}(d1)] and experiencing different inhomogeneities induced by the fabrication process, which randomizes the MCD value. The average of CE sites only [grey data in Fig.~\ref{fig:EdgeCompacton}(e)] shows an oscillation around zero, as these two sites support mainly the FBs that bound the light on a single unit cell. For $\Phi=\pi$, the nontriviality of the lattice becomes evident with MCD values around $-0.5$. Indeed, the MCD is closest to $-0.5$ around the wavelength of $640$ nm, for which the nontriviality condition is fully fulfilled. It is proportional to an integer value of Zak-phase $Z_p[\pi]=2MCD=\pm 1$. Therefore, we have experimentally demonstrated the trivial and nontrivial properties of the graphene-like ribbon and showed directly the bulk-boundary correspondence for the topologically compact edge states. 
\begin{figure}[]
\includegraphics[width=\textwidth]{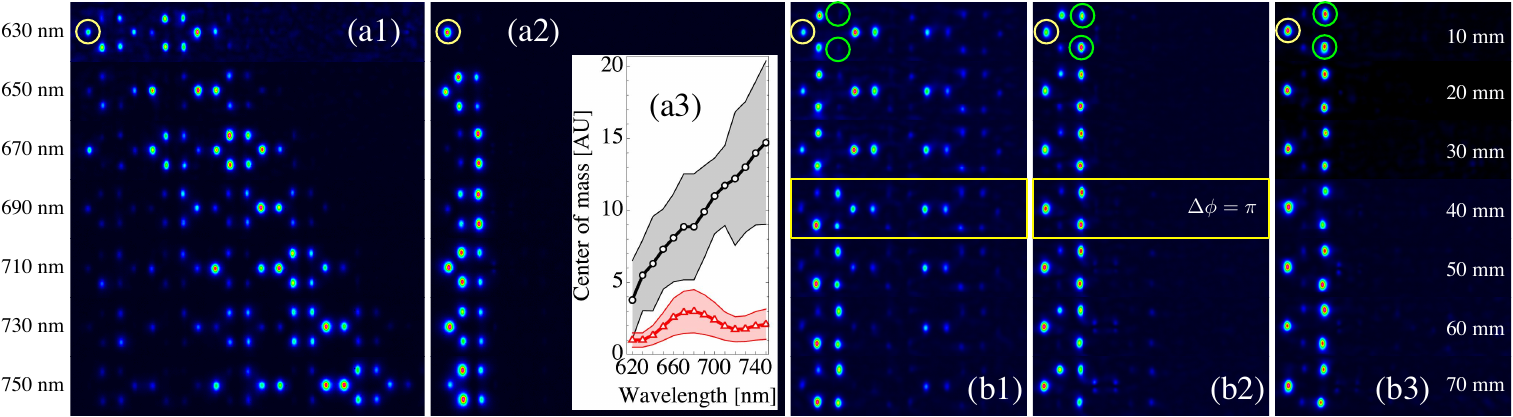}\
  \caption{(a) Edge $A$ site excitation: (a1) and (a2) output intensities for the excitation of S and SP lattices, respectively, and for the wavelengths indicated to the left. (a3) Center of mass versus wavelength for the S (black) and SP (red) lattices, with the shaded area showing the second moment. (b) SLM edge excitation: (b1) and (b2) sweep of the relative input phase $\Delta\phi$ in between site $A$ (yellow) and sites $C,E$ (green). Vertical images show different input phases around $\Delta\phi=\pi$ (yellow frame). (b3) Excitation of the edge compact topological state at $\Delta\phi=\pi$ versus the propagation distance $z_s$ indicated to the right.} 
\label{fig:Robustness}
\end{figure}

To study the edge excitation, we first excite the site $A$ at the left edge of both lattices. In the trivial $\Phi=0$ lattice, we observe only transport with the light strongly repelled from the edge; see Fig.~\ref{fig:Robustness}(a1). It is a direct consequence of the absence of edge states from the spectrum shown in Fig.~\ref{fig:Spectrum}(c). In the presence of nontrivial flux, we observe a remarkable effect. The light remains trapped to the first unit cell, oscillating locally due to the simultaneous excitation of three edge modes. In Fig.~\ref{fig:Robustness}(a2), a revival is observed around $720$ nm. To compare the light propagation through these lattices, we plot the centre of mass as a function of wavelength, with a shaded area showing the second moment as an indicator of the profile width. We observe that, in the trivial lattice (black data), the light escapes away from the edge and strongly disperses. In the topological system (red data), the light oscillates within the first cell with almost no dispersion, confirming the observed revival. While the tuning condition was optimized in the fabrication around $640$ nm, the results show that the destructive interference condition at $\Phi=\pi$ is valid beyond the tuning zero regime. Nevertheless, we stop measuring at $750$ nm to avoid next-nearest-neighbour coupling effects.

We further investigate the robustness of the compact zero-edge modes. An image generator simultaneously excites different lattice sites with a given phase structure. We generate a three-site input state at $640$ nm and excite sites $A$, $C$ and $E$ at the left edge. As shown by yellow and green circles in Fig.~\ref{fig:Robustness}(b), $C$ and $E$ are excited in phase, while the excitation phase in $A$ is varied with a relative phase difference around $\Delta\phi=\pi$. The results for trivial lattice in Fig.~\ref{fig:Robustness}(b1) show clear signatures of dispersion and FB oscillations without a tendency for edge localization. On the other hand, for $\Phi=\pi$ [see Fig.~\ref{fig:Robustness}(b2)], we observe that the light gets trapped at the first unit cell with almost no dispersion to the lattice bulk. The localization persists regardless of the phase structure of the input light, thus showing the robustness of the compact edge modes to input phase deviations. This is particularly important in applications in which the phase of an excitation beam is subject to fluctuations.

Finally, we study the compact-edge-mode propagation stability and robustness to fabrication tolerances. We fix the input phase to $\Delta\phi=\pi$ and fabricate a z-scan configuration~\cite{spirodrigo}, in which the excited sites $A$, $C$ and $E$, at the first unit cell only, have the maximum length allowed by the substrate (70 mm), while the rest of the lattice sites have a shorter $z_s$ length. To follow the state evolution along the lattice, we fabricate $14$ lattices with different $z_s$ and show a selected set of output profiles in Fig.~\ref{fig:Robustness}(b3). We observe an excellent stability of the compact edge mode along propagation. As each lattice is fabricated independently of the others, a random fabrication disorder is present in every sample, which allows us to conclude that the zero-edge mode is resistant to the small variations in the lattice structure. This result is also confirmed numerically by a deliberate introduction of disorder into the Hamiltonian (\ref{eq:Hamiltonian}) (Supplementary information).

\section{DISCUSSION}\label{sec:Discussion}

The main achievements of the presented study are the generation of highly robust topologically protected compact edge modes and the establishment of a novel procedure for the assessment of topological nontriviality via the mean chiral displacement. Demonstration of the stable compact edge modes is a decisive stepping stone towards unidirectional topological insulators. Further upgrades, including the non-Hermitian Hamiltonian to achieve unidirectional mode propagation and the inclusion of higher dimensions, will make the topological photonic lattices a faithful model for ultra-low energy electronics or photonics computing. The experimental implementation of synthetic flux and the corresponding MCD estimation open the door to the detection of topological nontriviality in multiband dispersionless systems. 

\section{\bf METHODS }

\vskip 1cm
\noindent{\bf The mean chirality displacement}
\vskip 1cm

The MCD is used to determine the Zak phase in 1D multi-band bipartite lattices with inherent chiral symmetry and isolated bands~\cite{chiral1,chiral2,chiral3}. It is an extension of the mean field displacement (MFD) method widely used in 1D two-band systems \cite{longhi,longhiMFD,SSHdiamond}, which is based on the geometric (Berry) phase. The geometric phase is proportional to the displacement of the centre of mass of Wannier functions, which form the basis complementary to the Bloch basis \cite{book}. 

In the literature, the MFD is determined numerically and experimentally by the mean displacement of the centre of propagating wavepacket initiated by single site lattice excitation \cite{longhi,longhiMFD,SSHdiamond}: 
\begin{equation}
M_{FD}(L_z)=\frac{1}{L_z} \int_{z=0}^{z=L_z}  \frac{\sum_{n=0}^{N-1}(x_n(z)-x_n(0)|\psi_n(z)|^2}{\sum_{n=0}^{N-1}|\psi_n(z)|^2} \; dz,
\end{equation}
where $L_z$ is the propagation distance in the $z$-axis direction and $x$ is position in ribbon, and $\Psi=[\psi_1\, \psi_2...\psi_N]^T$ is the wavepacket.

 
In our numerical calculations and experiments, we excite the sites of the central unit cell hosting an eigenstate of the chiral operator, i.e., $A$, $C$, and $E$ for the state with the eigenvalue $1$ or $B$, $D$, and $F$ for the state with the eigenvalue $-1$. The sites are excited one by one and the MCF is determined as an average value of the three displacements. 

Finally, the Zak phase is obtained as a twofold value of the mean chiral displacement
$Z_p[\pi]=2M_{CD}$ and, according to the bulk-edge correspondence, equals the number of topologically protected edge modes~\cite{topo_review1,topo_review2}.

The details of the calculation and chirality model are presented in Supplementary information.

\vskip 1cm
\noindent{\bf Lattice fabrication and characterisation}
\vskip 1cm

The lattices were fabricated by direct laser writing technique~\cite{alexfs} sketched in Fig.~\ref{fig:EdgeCompacton}(a). A femtosecond laser of $1030$ nm was tightly focused onto a borosilicate glass wafer, which was translated along $z$ to inscribe the waveguides and in $x-y$ plane to achieve the designed lattice profile. The synthetic flux was induced by replacing some $S$ lattice sites with dipole-like $P$ waveguides, as shown in Fig.~\ref{fig:EdgeCompacton}(c1) and (c2), and optimizing the interaction between them by a wavelength tuning scheme~\cite{spdimer,spirodrigo}. The zero-detuning condition was achieved at $640$ nm, but the flux $\Phi=\pi$ was obtained for a much broader wavelength range as demonstrated in the experiments.  

The characterization experiments were performed by a horizontally polarized supercontinuum (SC) laser source in the wavelength range of $\{630,750\}$ nm. When the excitation wavelength is increased, the guided-mode width
increases and, therefore, the coupling constant becomes larger~\cite{SSHdiamond}, effectively increasing the propagation length. Hence, by varying the input wavelength, we can study a system with a fixed glass length, $L$, but at a different dynamical instant. 

The edge mode profile was excited by an amplitude- and phase-modulated beam obtained from an image generator setup (see Supplementary information for details). The beam was first amplitude-modulated by a Holoeye LC2012 transmission spatial light modulator (SLM) to enable simultaneous excitation of multiple waveguides. Subsequently, the phase was modulated by a Holoeye PLUTO reflection SLM. The final image is transferred to the input facet of the photonic chip by a telescope and a $4 \times$ microscope objective. The array output is imaged using a $10 \times$ objective and recorded by a CCD camera.

\backmatter
\section*{Declarations}

\bmhead{Data avialiabilty}
All data needed to evaluate the conclusions in the paper are available 
in this Article or its Supplementary Information. The data that support 
the findings of this study are available from the corresponding authors 
upon reasonable request.

\bmhead{Acknowledgments}
This research was supported in part by Ministry of Science, Technological Development and Innovation of the Republic of Serbia (451-03-9/2021-14/ 200017), Millennium Science Initiative Program ICN17$\_$012, FONDECYT Grant 1231313, and the IRTG  ``Imaging Quantum Systems" No.~437567992 by the German Research Foundation.

\bmhead{Author contributions}
G.C.-A. performed analytical calculations; M.N. and G.G. designed numerical model; G.C.-A., M.N., and G.G. performed numerical simulations; P.V. and R.V. designed and performed the experiments; G.G., J.P., R.A.V. and A.M. analyzed results and performed checking of results; R.A.V. and A. M. initialized and supervised the project. All the authors contributed to the final form of the manuscript.

\bmhead{Competing interests}
The authors declare no competing interests.

\bmhead{Additional information}

\bmhead{Supplementary information}
The online version 
contains supplementary material available at ...

\bmhead{Correspondence and requests for materials} Address to R.A. Vicencio and A. Maluckov

\end{document}


\renewcommand{\theequation}{S\arabic{equation}}
\renewcommand{\thefigure}{S\arabic{figure}}
\renewcommand{\bibnumfmt}[1]{[S#1]}
\renewcommand{\citenumfont}[1]{S#1}


\title{Supplementary information for ``Observation of topologically protected compact edge states in flux-dressed photonic graphene ribbon''}

\author[1]{ Gabriel C\'aceres-Aravena}
\email{gabriel.caceres-aravena@uni-rostock.de}
\equalcont{These authors contributed equally to this work.}

\author[2]{ Milica Nedi\'c}
\email{milica.brankovic@vin.bg.ac.rs}
\equalcont{These authors contributed equally to this work.}

\author[3]{ Paloma Vildoso}
\email{palomavildoso1@gmail.com}
\equalcont{These authors contributed equally to this work.}

\author[2]{ Goran Gligori\'c}
\email{goran79@vin.bg.ac.rs}

\author[2]{ Jovana Petrovic}
\email{jovanap@vin.bg.ac.rs}

\author[3]{ Rodrigo A. Vicencio}
\email{rvicencio@uchile.cl}

\author*[2]{ Aleksandra Maluckov}
\email{sandram@vin.bg.ac.rs}

\affil[1]{\orgdiv{Institute of Physics}, \orgname{University of Rostock}, \city{Rostock}, \postcode{18051}, \country{Germany}}

\affil[2]{\orgdiv{Vin\v ca Institute of Nuclear Sciences, National Institute of the Republic of Serbia}, \orgname{University of Belgrade}, \orgaddress{\street{P.O.B. 522}, \city{Belgrade}, \postcode{11001},  \country{Serbia}}}

\affil[3]{\orgdiv{Departamento de Física and Millenium Institute for Research in Optics–MIRO}, \orgname{Facultad de Ciencias Físicas y Matemáticas, Universidad de Chile}, \orgaddress{\city{Santiago}, \postcode{8370448},  \country{Chile}}}

\date{\today}

\maketitle

\noindent{\bf {1. Localized modes in ribbon}}

\vskip 1cm

\noindent{\it Compact localized modes} 

The compact localized states (CLS) are steady states of the $\pi$-flux Hamiltonian. Analytically to derive the CLSs we start from model equation Eq.~(2) (main text) transformed to the configuration (lattice) space
\begin{eqnarray}
i\frac{da_{n}}{dz} & =& b_{n}Ve^{-i\phi}+f_{n}Ve^{i\phi}+D_{n-1}V \nonumber\\
i\frac{db_{n}}{dz} & =& a_{n}Ve^{i\phi}+c_{n}Ve^{-i\phi} \nonumber \\
i\frac{dc_{n}}{dz} & =& b_{n}Ve^{i\phi}+d_{n}Ve^{-i\phi} \nonumber\\
i\frac{dd_{n}}{dz} & =& c_{n}Ve^{i\phi}+e_{n}Ve^{-i\phi}+A_{n+1} \\
i\frac{de_{n}}{dz} & =& d_{n}Ve^{i\phi}+f_{n}Ve^{-i\phi} \nonumber\\
i\frac{df_{n}}{dz} & =& a_{n}Ve^{-i\phi}+e_{n}Ve^{i\phi} \nonumber
\label{jednacine}
\end{eqnarray}
where $n$ is the index of unit cell, $\psi_n=(a\, b\, c\, d\,e\, f)_n$ is the six-component vector, $V$ is the coupling strength (Fig. 1 in the main text) and $\phi=\Phi/6$. The presence of flux is modeled by complex coupling coefficients inside the unit-cells.

Being a steady state of $\Phi=0$ and $\Phi=\pi$-flux ribbon, the CLS can be presented in a form $(a_n, b_n, c_n, d_n,e_n,f_n)(z)=exp(-i\lambda z)(A_{n},B_{n},C_{n},D_{n} ,E_{n},F_{n})$. Therefore the set of eigenvalue equations ($V=1$) is:
\begin{equation}
\begin{aligned}
\lambda A_{n} & = B_{n}e^{-i\phi}+F_{n}e^{i\phi}+D_{n-1} \\
\lambda B_{n} & = A_{n}e^{i\phi}+C_{n}e^{-i\phi} \\
\lambda C_{n} & = B_{n}e^{i\phi}+D_{n}e^{-i\phi} \\
\lambda D_{n} & = C_{n}e^{i\phi}+E_{n}e^{-i\phi}+A_{n+1} \\
\lambda E_{n} & = D_{n}e^{i\phi}+F_{n}e^{-i\phi} \\
\lambda F_{n} & = A_{n}e^{-i\phi}+E_{n}e^{i\phi} \\
\label{jednacine}
\end{aligned}
\end{equation}
where $\lambda$ is eigenenergy. 

\begin{figure}
  \center{\includegraphics[width=8cm]{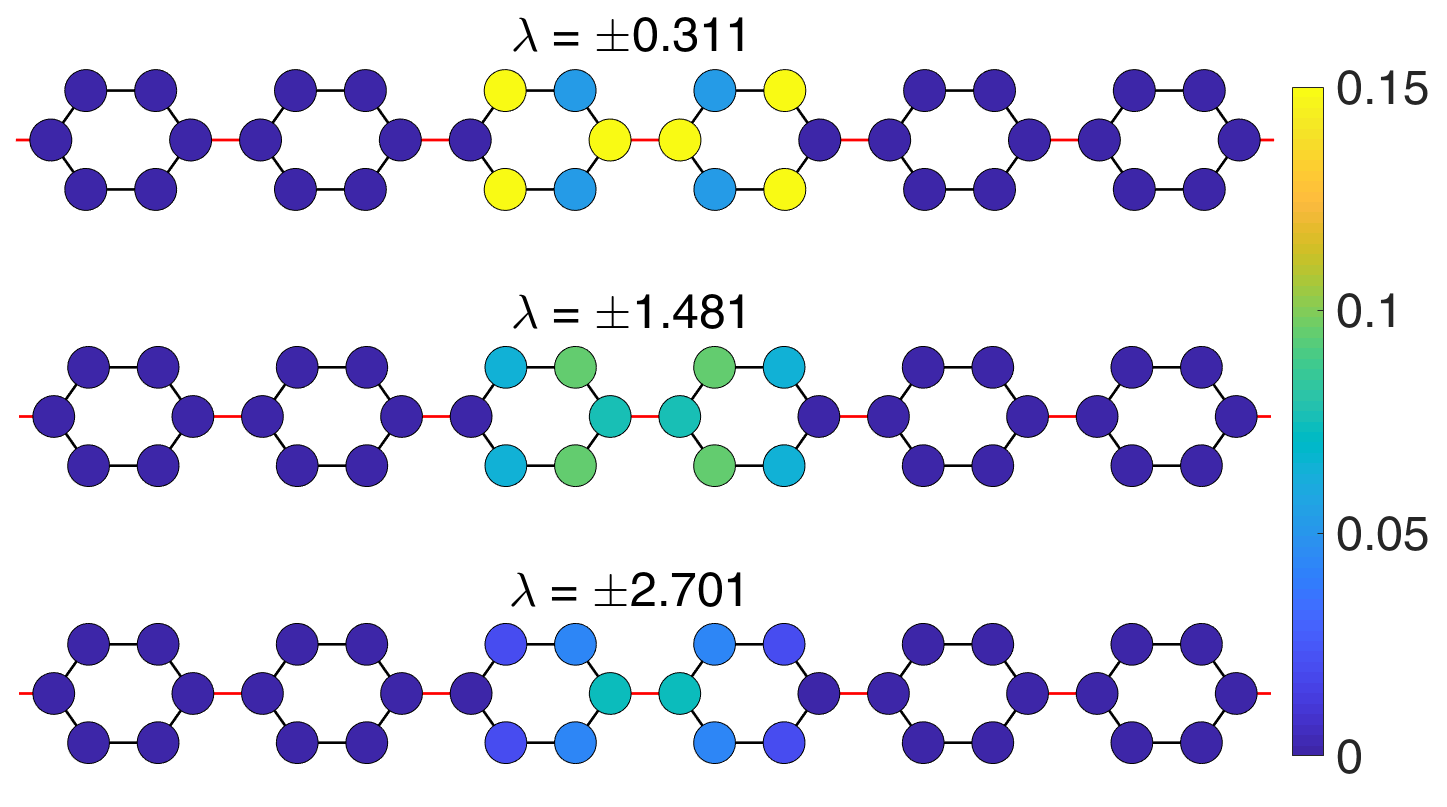}}
\caption{The fundamental compact localized modes in $\pi$-flux graphene ribbon. The CLSs corresponding to the same absolute value of $\lambda$ differ by phase distribution among the nonzero amplitude sites, Eqs. \ref{cls}. } 
\label{fcls}
\end{figure}

After simple algebraic procedure the CLS amplitudes can be estimated from expressions:
\begin{equation}\label{cls}
\begin{aligned}
\{A, \,B,\,C,\,D,\,E,\, F\}_n & =\kappa\{0,\,
1,\,-\lambda e^{i\phi},\,(\lambda^2-1)e^{2i\phi},\,-\lambda e^{3i\phi},\, e^{4i\phi} \}\\
\{A, \,B,\,C,\,D,\,E,\, F\}_{n+1} &=
\frac{\kappa (3-\lambda^2)}{1-\lambda^2}\{\lambda(1-\lambda^2)e^{2i\phi},\,\lambda^2 e^{3i\phi},\,
-\lambda e^{4i\phi},\,0,\lambda,\,
\lambda^2e^{i\phi}     \}   
\end{aligned}
\end{equation}
where $\kappa$ is an arbitrary constant ($\kappa=1$ due to convenience). Two FBs at fluxless ribbon are singular due to the crossing of FBs with dispersion bands, while at $\Phi=\pi$ the spectrum consists of six isolated  FBs. The fundamental CLSs hosted by two unit-cells, in the last case form the  compact nonorthogonal basis (Fig. \ref{fcls}).

\begin{figure}\centering
\includegraphics[width=10cm]{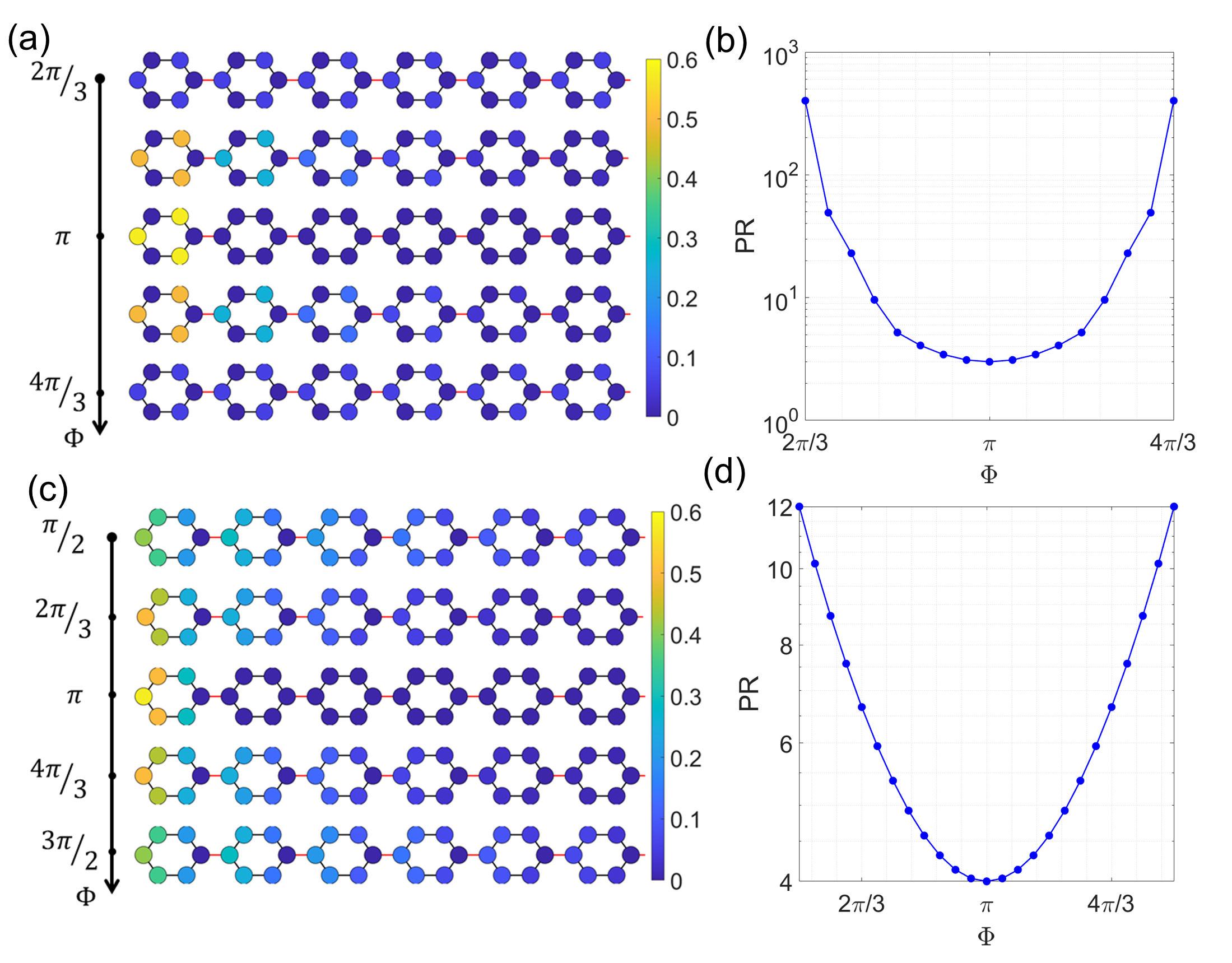}
  \caption{ (a) The schematic view of the amplitude distribution of zero-edge double degenerated mode ($\lambda=0$) and (c) double degenerated edge mode ($\lambda=\sqrt{3}$) in graphene ribbon vs. flux in the area $\Phi=[2\pi/3,4\pi/3]$. (b) and (d) The participation ratio in the logarithmic scale for modes shown in (a) and (c), respectively.  } 
\label{edgezero}
\end{figure}

\vskip 0.5cm
\noindent{\it Edge modes} 
\vskip 0.5cm
The edge modes are caged to the lattice left/right edge, Figure \ref{edgezero} (a) and (c). Their topological properties are determined in this study.  
The edge mode amplitudes can be analytically estimated from recursive relations obtained from Eq. (1) for $\lambda=0,\pm\sqrt{3}$ assuming the $D$ sites (left ribbon edge) or $A$ sites (right ribbon edge) are empty. Here are presented expressions for the left edge mode:
\begin{eqnarray}
\{A, \,B,\,C,\,D,\,E,\, F\}_{n} & =&A_n\left\{1,\,\frac{\lambda\,e^{i\phi}}{(1-\lambda^2)} ,\,\frac{e^{i2\phi}}{\lambda^2-1} ,\,0,\,\frac{e^{-i2\phi}}{\lambda^2-1} ,\, \frac{\lambda\,e^{-i\phi}}{1-\lambda^2}  \right\}\nonumber\\
A_{n+1} & =&- \frac{2\cos{(3\phi)}}{\lambda^{2}-1}A_n.
\end{eqnarray}
Taking arbitrary value of $A_1=c$ we obtain $A_{n\ne 1}=(-2\cos(3\phi)/(\lambda^2-1))^{n-1}c$, which in the finite lattice case with $N$-sites provides condition:
\begin{equation}
\left(-\frac{2\cos(3\phi)}{\lambda^2-1}\right)^N c=0
\label{6}\ .
\end{equation}   
Only in the fully FB ribbon, $\Phi=\pi$ ($\phi= \pi/6$) Eq. (\ref{6}) is satisfied for all $N$, while in the other flux configurations the edge modes posses $N$-dependent tails, as confirmed by the calculation of  the participation ratio:
\begin{equation} 
PR = \frac{\mathcal{P}^2}{2N} \sum_{i}^{N} \left( |\Psi_i|^4 \right)^{-1},\quad \Psi_i=(A\, B\, C\, D\, E\, F)_{i}^T\nonumber
\end{equation}
where $\mathcal{P} = \sum_{i} \Psi_i^{\dagger}\Psi_{i}$ is the total power, Fig. \ref{edgezero} (b) and (d). The participation ratio gives information regarding the fraction of strongly excited lattice sites. 
Dynamical simulations confirm extreme robustness of edge modes.

\vskip 1cm

\noindent{\bf {2. Symmetries of armchair graphene-like ribbon}}
\vskip 1cm

Following the Altland-Zirnbauer symmetry classification of 1D systems, the graphene armchair ribbon belongs to AIII symmetry class \cite{altmland,zero}. 
It is characterized by bipartite symmetry regardless of the flux value. As a consequence, the energy band spectrum is symmetric with respect to zero energy.

The bipartite symmetric Hamiltonian $\hat{H}$ is related to a chirality operator $\hat{G}$, 
$\hat{H}=\hat{G}\hat{H}\hat{G}=-\hat{H}$, which can be expressed in a symbolic form:
\begin{equation}
\quad \hat{G}=\hat B + \hat D + \hat F - ( \hat A + \hat C + \hat E )\ ,
\end{equation}
with $\hat A = \sum_n \ket{A_n}\bra{A_n}$ being the sublattice projector for the $A$ sites, $\hat B$ for $B$ sites, etc.  Chirality operator provides the possibility to create the topologically non-trivial phases in the system. 

In addition, the armchair fluxless ribbon \cite{bauer-1,bauer-2} is symmetric with respect to the horizontal axis for any number of hexagons and with respect to the vertical axis for even number of hexagons. The introduction of nonzero flux asymmetrizes the ribbon, which is associated with opening of gaps between the periphery bands at the boundaries of the first Brillouin zone (BZ). 

The gap between the central bands is closed and reopened at the flux values $\Phi=2\pi/3$ and $\Phi=4\pi/3$ at the center of first BZ. At these particular flux values, the C3 symmetry inside of unit cells (hexagons) is reestablished. Hence, the opening/closing of the central gap, which is followed by the creation/destruction of the chiral zero-edge states, can be related to establishing/breaking of this symmetry. For $\Phi=\pi$ zero edge mode is hosted by only three sites inside the boundary unit cell.

\vskip 1cm
\noindent{\bf {3. Determination of topological phase: bulk-edge correspondence}}
\vskip 1cm
The nontrivial topological phase of periodic systems (lattices, chains) is described by quantized topological invariant derived from Berry (geometric) phase after the adiabatic evolution of the system \cite{book,azboth}. The role of the topological invariant in 1D lattice is played by the winding number or related to it, Zak-phase, which describes how a particular function (connected with Bloch states) winds around the 1st Brillouin zone. These two quantities in 1D are proportional: $Z_p=W\pi$, where $W$ is winding number and $Z_p$ the Zak phase.  
In 1D lattices integer values of Zak-phase indicate the presence of topologically protected edge states which is in pursuance of the bulk-edge correspondence.

The SSH lattice is a paradigm of 1D topological insulators \cite{} formed of two-sites' unit cells. Peculiarity of the SSH lattice is topological nontriviallity in the parametric area $v/w>1$ described by $Z_p=\pm 1$, where parameters $v,w$ are the intra and inter site coupling strengths, respectively. Strictly, the threshold value is dependent on the number of cells. The band spectrum is formed of two bands which are connected by the topologically protected edge (zero) mode in topologically nontrivial phase. There are a few mathematical methods to derive the value of $Z_p$, the geometrical, mean chiral displacement and projector method. 

The first attempts regarding the topology of multi-band periodic systems are developed on the effective SSH-like equivalent two-band systems \cite{multissh}. Analyzing our multi FB system we  theoretically and experimentally prove the nontrivial topological properties of central bands and extremely robust compact zero-edge modes. Moreover, in the flux range $[2\pi/3,4\pi/3]$, the topological non-triviality is numerically confirmed too. 
The topological nontriviality in the rest of parametric space of flux-dressed ribbon (see the main text) is just indicated by utilizing the projector method. Final confirmation of these finding requires additional study because the other two methods presently could not be applied to periphery bands.

\vskip 0.5cm
\noindent{\it The geometric method: reduction to SSH-like Hamiltonian}
\vskip 0.5cm
The governing idea of the geometric method is based on the reduction of the Hamiltonian to the SSH-type (two-band) one, which can be expressed as:
\begin{equation}
\label{reduction}
\hat{H}=\left(
\begin{array}{cc}
 \hat{0} & \hat{h}  \\
 \hat{h}^\dagger& \hat{0}  \\
 \end{array}
\right).
\end{equation}
In the $N$-dimensional SSH Hamiltonian the block matrices $\hat{h}$ are $(N/2)\times (N/2)$ matrices. 

\begin{figure}
   \centering
    \includegraphics[width=0.95\textwidth]{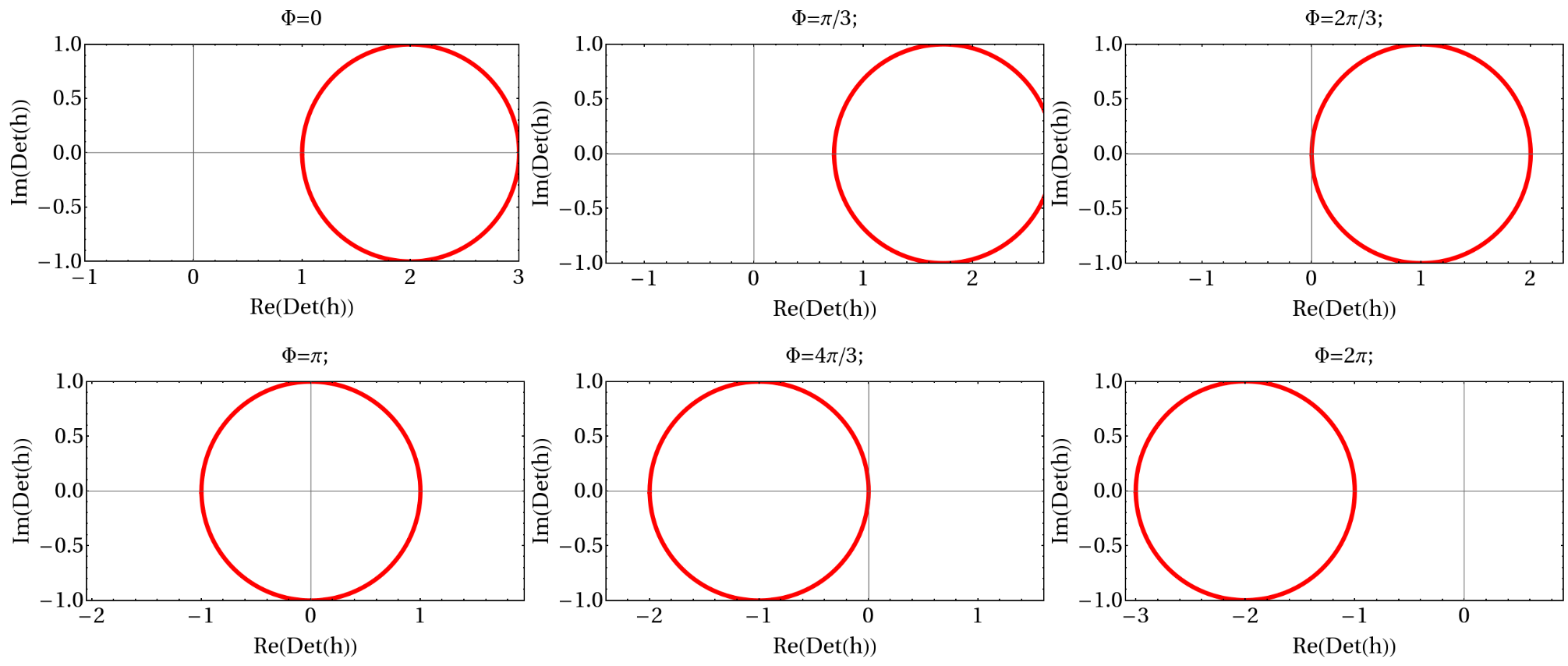}
    \caption{The determinant of $h$ from Eq.(\ref{subspaceofH}) for $\Phi=\{0,\pi/3,2\pi/3,\pi,4\pi/3,2\pi\}$.} \label{krug}
\end{figure}

We found that the following unitary matrix $S$ ($S^\dagger S = I$, therefore $S^{-1}=S^\dagger$):
\begin{align}
S=\left(
\begin{array}{cccccc}
 1 & 0 & 0 & 0 & 0 & 0 \\
 0 & 0 & 0 & 0 & 0 & 1 \\
 0 & 1 & 0 & 0 & 0 & 0 \\
 0 & 0 & 0 & 1 & 0 & 0 \\
 0 & 0 & 1 & 0 & 0 & 0 \\
 0 & 0 & 0 & 0 & 1 & 0 \\
\end{array}
\right)\ ,
\end{align}
transforms $H(k)$ (Eq. (2) in main text) into :
\begin{align}
S^{-1} H S=\left(
\begin{array}{cccccc}
 0 & 0 & 0 & e^{-i k_x} & e^{i \phi } & e^{-i \phi } \\
 0 & 0 & 0 & e^{-i \phi } & 0 & e^{i \phi} \\
 0 & 0 & 0 & e^{i \phi } & e^{-i \phi } & 0 \\
 e^{i k_x} & e^{i \phi } & e^{-i \phi } & 0 & 0 & 0 \\
 e^{-i \phi } & 0 & e^{i \phi } & 0 & 0 & 0 \\
 e^{i \phi } & e^{-i \phi } & 0 & 0 & 0 & 0 \\
\end{array}
\right)\ ,
\end{align}
which can be rewritten in the form  (\ref{reduction}) in terms of off-diagonal matrix $h$ 
\begin{align} \label{subspaceofH}
h=\left(
\begin{array}{ccc}
 e^{-i k_x} & e^{i \phi } & e^{-i \phi } \\
 e^{-i \phi } & 0 & e^{i \phi } \\
 e^{i \phi } & e^{-i \phi } & 0 \\
\end{array}
\right).
\end{align}
 By calculating the determinant of the off-diagonal sub-matrix we obtained $\det(h)=-\cos k_x+2\cos(\Phi/2)+i\sin k_x$, where $\Phi=6\phi$. 
 In the complex plane this determinant can be presented by unit circle which position depends on the parameter $\Phi$. The winding number can be defined by how many types loops encircle the origin. Here the origin is circled ones for the fluxes $\Phi = [2\pi/3,4\pi/3]$. From this follows that only in this parameter region the winding number is $1$, Fig. \ref{krug}.

\begin{figure}
   \center{
    \includegraphics[width=12cm]{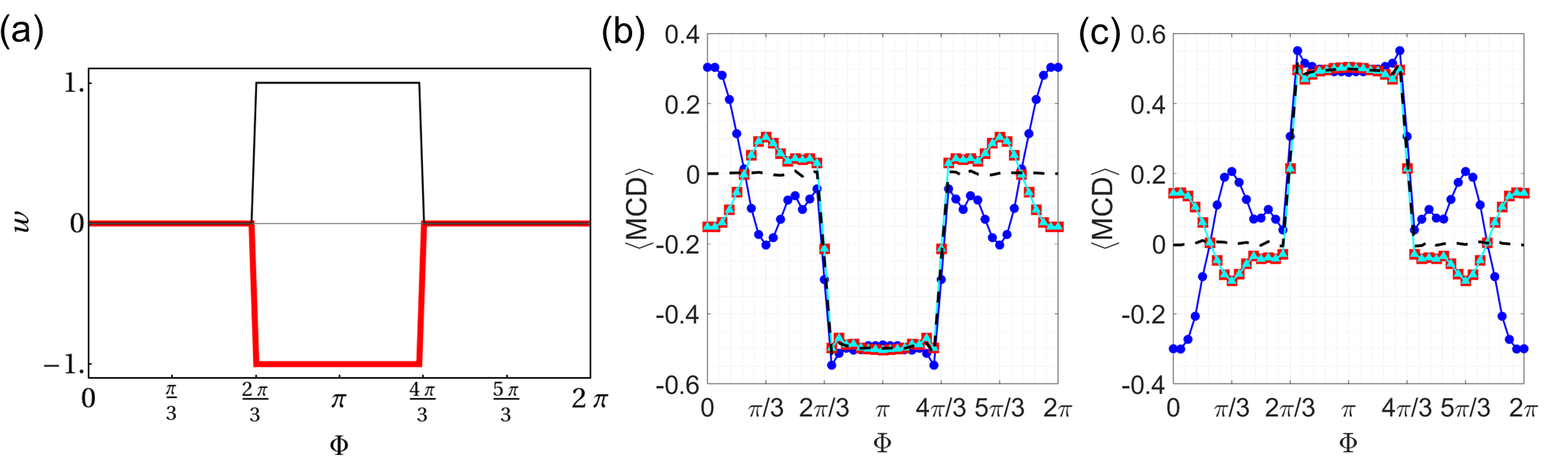}
    \caption{(a) The calculation of the winding number using equation (\ref{wind}) with $h$ for the red line and $h^\dagger$ the black line. (b) Numerically obtained values of MCD by initially exciting single A (blue circle), C (red squares), and E (cyan triangles) site belonging to the first sublattice. (c) The same as (b) just by initially exciting single D (blue circle), B (red squares), and F (cyan triangles) site belonging to the second sublattice. The black dashed lines are the average values of the MCD single-site excitations in the corresponding sublattice.}} \label{wind_num}
\end{figure}

In addition, the value of the winding number $w$ can be determined from expression:
\begin{align}\label{wind}
    w = \frac{1}{2\pi i} \int_0^{2\pi} dk_x \; \frac{\partial}{\partial k_x} \log\left( \det(h(k_x)) \right),
\end{align}
showing that it is unity in the range of fluxes $\Phi=[2\pi/3,4\pi/3]$, Fig. \ref{wind_num}. Thus, the topological nontriviallity of the central bands and via the bulk-edge correspondence of the compact edge modes is confirmed.

\vskip 0.5cm
\noindent{\it The mean field/chirality displacement}
\vskip 0.5cm

In the root of the Mean Field Displacement (MFD) method and the projector methods is the concept of geometric, i.e. the Berry phase which describes how a global phase accumulates as a complex Bloch eigenvector is carried around a closed loop in the $\vec{k}$ space inside the 1st Brillouin zone. In the 1D periodic system in the basis of Wannier functions, the Fourier-transform partners to the Bloch functions, the Berry phase is proportional to the displacement of the center of mass of basis vectors \cite{book}. In the literature, the MFD method has been widely used in the 1D two-band systems to derive numerically and experimentally the value of the topological invariant 
of isolated band, which is determined by the mean displacement of the center of propagating wavepacket initiated by single site lattice excitation \cite{longhi,longhiMFD,SSHdiamond}: 
\begin{equation}
M_{FD}(L_z)=\frac{1}{L_z} \int_{z=0}^{z=L_z} \overline{P}(z) \; dz=\frac{1}{L_z} \int_{z=0}^{z=L_z}  \frac{\sum_{n=0}^{N-1}(x_n(z)-x_n(0))|\psi_n(z)|^2}{\sum_{n=0}^{N-1}|\psi_n(z)|^2} \; dz,
\end{equation}
where $L_z$ is the propagation distance in the $z$-axis direction and $x$ is position in ribbon, and $\Psi=[\psi_1\, \psi_2...\psi_N]^T$ is the wavepacket. 
The Zak-phase is given by the expression:
$Z_p[\pi]=2\pi M_{FD}$. 
 The integer value of $Z_p$ in units of $\pi$ is related with the total number of edge states formed in the gaps between the isolated bands, just after the topological phase transition. This is a confirmation of the bulk-edge correspondence \cite{topo_review1,topo_review2}.

The MFD method is experimentally utilized to clarify the theoretical predictions in diverse 1D and quasi 1D lattices \cite{longhi,longhiMFD,SSHdiamond}. 
It has been proved that the value $2M_{FD}$ approaches asymptotically the winding number \cite{longhiMFD} (Fig. \ref{wind_num}), or related Zak phase (with a $\pi$ factor) \cite{longhi} for large values of the propagation distance. 

Strictly speaking the above described procedure is fully developed for the 1D lattice systems with two bands \cite{chiral1,longhi,longhiMFD,SSHdiamond}. However, the MFD approach can be applied to the lattices' with multilevel spectra which geometry and related symmetries allow the reduction to the effective two-level spectrum \cite{chiral2,chiral3,multissh}. In other circumstances problem is model-dependent.

Further implementations of this method for more complex lattices had been made taking advantage of the chiral symmetry of the bipartite lattices \cite{chiral1,chiral2,chiral3}. 
The main idea is to utilize the sublattice related peculiarities of such systems. Thus in the chiral lattice with $4$ sites ($2$ from each sublattice)  \cite{chiral2}, the MFD is calculated using two different initial excitations of sites within the same unit cell, each meaning the excitation of just one of sublattices, are followed. The corresponding MFD are summed and their average value is classified as a Mean Chiral Displacement (MCD)\cite{chiral1}, Fig. \ref{wind_num} (b). The winding number (Zak-phase) is then $Z_p=2MCD/\pi$. Here we adopted the same idea to our chiral ribbon which unit cell consists of six sites, three from each of sublattice. Chiral symmetry is associted with the chirality operator $G$. We excite each of sites inside the middle unit cell hosting the eigen-states of the chiral operator, $A,C,E$ or $B,D,F$ (the eigenvalue of $G$ are respectivelly, $1$ and $-1$). The averaged value of three MFD is taken to determine the Zak-phase of central ribbon bands. The shape of the MCD vs. $\Phi$ fully corresponds to the $W$ vs. $\Phi$ derived by geometric method, Fig. \ref{wind_num}.

\vskip 0.5cm
\noindent{\it Calculation of Zak-phase: projector method}

\vskip 0.5cm
The projector method is established in the basis of Bloch vectors \cite{book, leaky}. The central feature is the projection operator of eigenmodes $\hat{P}(k)=|\tilde{\psi} (k)\rangle \langle \tilde{\psi} (k)|$ \cite{phothalleffect,book,leaky} from which the topological invariant is derived:
\begin{equation}\label{eqq}
Zp=\Im \left( \ln ( \Tr \left[ \prod_{n=1}^{N}\hat{P} (k_n ) \right] ) \right)\ ,
\end{equation}
where $k_n=2\pi n/L$ is the discrete momentum. 

The numerical procedure starts by solving the eigenvalue problem of the ribbon Hamiltonian in $k$-space, which is now discretized in $N$-grids separated by $2\pi/N$ distance.  For each $k_i, \, i=[1,..,N]$ within the 1st BZ, $k=[-\pi,\pi]$ the set of $6$ eigenvalues and corresponding eigenvectors is determined. Each of $6$ eigenvalues belongs to different bands. The $6 \times 6$  projector matrix is formed as the correlation matrix between the eigenvectors from the whole eigenvalue set. After straightforward numerical calculation the values of Zak-phase determined by Eq. \ref{eqq}, for each band are summarized in Table I.

\begin{table}\centering
	\begin{tabular}{||c c c c c c c||} 
		\hline
		Flux & band 1 & band 2 & band 3 & band 4 & band 5 & band6 \\ [0.5ex] 
		\hline\hline
		$\Phi=0$ & 0 & 0 & 0 & 0 & 0 & 0 \\ 
		\hline
		$0<\Phi<2\pi/3$ & -1 & 1 & 0 & 0 & 1 & 1 \\ 
		\hline
		$2\pi/3<\Phi<4\pi/3$ & 1 & -1 & 1 & -1 & -1 & 1 \\ 
		\hline
		$4\pi/3<\Phi<2\pi$ & 1 & -1 & 0 & 0 & -1 & 1 \\ 
		\hline
		$\Phi=2\pi$ & 0 & 0 & 0 & 0 & 0 & 0 \\  
			\hline
 
	\end{tabular}
    \caption{Table I: The values of Zak -phase for each band for different values of flux.}
    \end{table}

In contrast to other methods the projector method gives information about the topological phase of all isolated bands, the central and periphery ones. Assuming the validity of the bulk-surface correspondence we claim that the edge modes between the nontrivial bands are nontrivial too, i.e. topologically protected modes.

\vskip 0.5cm
\noindent{\it Theoretical vs Experimental model}
\vskip 0.5cm

The theoretical model is composed of $6$ sites per unit cell as described in the main text and at the beginning of the supplementary material. On the other hand, the experimental model for $\pi$ flux is composed of $12$ sites per unit cell but can be mapped to the theoretical model. The mapping of the experimental model without flux is simple given by assuming the sites are waveguides with the same propagation constant. 
However, in the case of the $\pi$ flux, the experimental model is composed of $12$ sites per unit cell and the value of the phase in the plaquette is achieved by a synthetic magnetic flux \cite{spirodrigo} owing a negative hopping from the second order photonic mode in a middle waveguide between two hexagons. 

\begin{figure}
    \centering
    \includegraphics[width=0.8\textwidth]{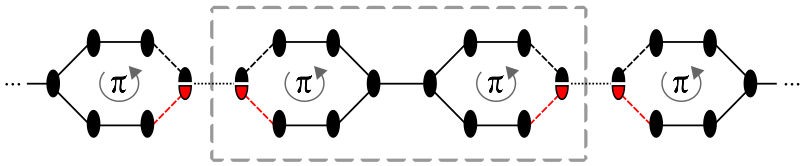}
    \caption{The scheme of the experimentally realized $\pi$-flux graphene ribbon. The gray dashed square denotes the $n$-th unit cell composed of 12 sites.} 
    \label{fig:exp_model}
\end{figure}

The Fig. \ref{fig:exp_model} shows the experimental model. 
When the couplings are all the same and the propagation constant (on-site energy) are equal to zero, then the model has 6 bands each with degeneracy 2. 
Given that each eigenvalue has degeneracy two, the system is reducible. The reduction is possible due to the left plaquette is the same as the right plaquette but mirrored. This mirroring is associated to a symmetry in the system that creates a degeneracy, therefore we can reduce the system to the half without losing information about eigenstates. 
In this manner, the 12 sites per unit cell system is equivalent to a 6 sites system after mirroring every other unit cell. Moreover, we calculate the eigenvalues of the momentum Hamiltonian (the basis of Bloch waves) and we find 12 values, each with degeneracy 2, therefore different 6 eigenvalues are $\{\pm 2.17,\pm 1.48,\pm 0.31\}$ which coincides with the 6 eigenvalues of the theoretical model when $\Phi=\pi$ and $V=1$ (main text).

\newpage 
\vskip 1cm
\noindent{\bf {4. Experimental characterization setups}}
\vskip 0.5cm

\begin{figure}[htbp]
\centering\includegraphics[width=0.6\textwidth]{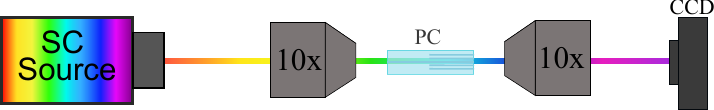}
\caption{Scheme of SC setup. It consists on a Supercontinuum (SC) laser source, a $10\times$ Microscope Objectives, and a beam profiler (mirrors, nd-filters, iris, polarizers, XYZ stages, goniometers are not including here for simplicity).}
\label{setupSC}
\end{figure}

{\it Supercontinuum characterization Setup}. The excitation of a photonic lattice is typically achieved by focusing a polarized gaussian-like laser beam at the input facet of a photonic lattice under study. Then, the output facet is imaged and obtained the intensity output profile on a CCD camera. This standard method is relatively simple to implement and it is limited to the excitation of a single waveguide only, at a fixed wavelength. The coupling constants of a photonic lattice depend on the system's geometry (determined by waveguide distances), but also on the excitation wavelength. The width of the mode profile at each waveguide increases for a larger wavelength and, therefore, for a fixed geometry it is possible to change the coupling constants by using a different excitation wavelength. This is the basis of the Supercontinuum setup, where by changing the input wavelength we modify the coupling constants, what  simultaneously alters the effective propagation distance of the system. The dynamical variable on tight-binding-like models is given by $``V z''$, where $z$ is the propagation coordinate and $V$ the coupling constant. For a fixed glass length, the coordinate $z$ is also fixed and, therefore, a modification of the coupling constants produces a shorter/larger effective dynamics. The experimental setup is sketched in Fig.~\ref{setupSC} and it relies on a Supercontinuum laser source (SC) YSL SC-5, emitting in the wavelength range of $450-2400$ nm. The operational range is narrowed down to $450-1450$ nm by using acoustic filtering, with a resolution of $5$ nm and $\sim 1$ mW at each wavelength. The SC beam is first horizontally polarized and, then, focused onto the input facet of the photonic chip (PC) using a $10 \times$ microscope objective. The output image is obtained by using a second $10 \times$ microscope objective and a beam profiler.


\begin{figure}[htbp]
\centering\includegraphics[width=0.85\textwidth]{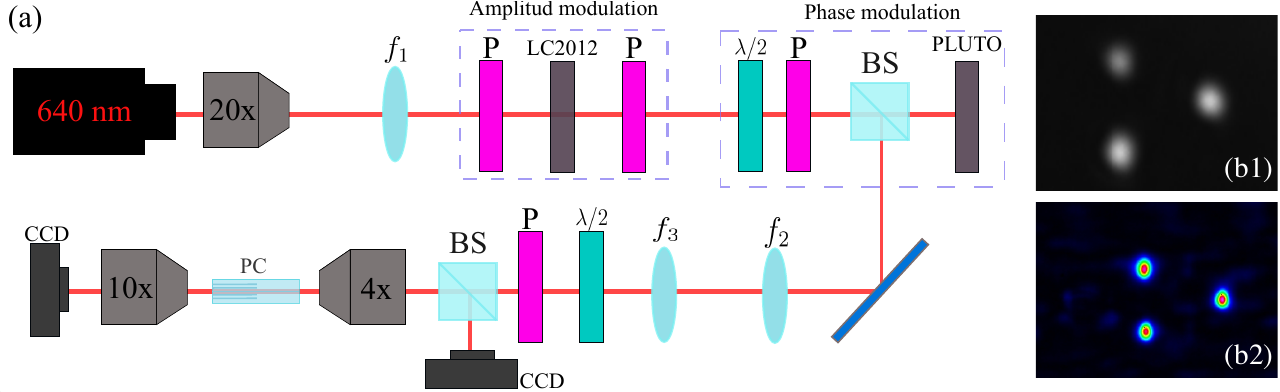}
\caption{Image setup. (a) Scheme of experimental image setup based on two SLM's to generate arbitrary input conditions. It consists on several linear polarizers (P), $\lambda/2$ waveplates, beam splitters (BS), several lenses, and two CCD cameras. (b1) Amplitude and phase modulation. (b2) Imaged output profile.}
\label{setupimg}
\end{figure}

\vskip 0.5cm
{\it Image Setup}. The excitation of compact edge states on a graphene ribbon demands the generation of an input condition having a specific amplitude and phase profile. To achieve this we use an image setup described in Figure~\ref{setupimg}(a). We first expand and collimate a $640$ nm laser beam using a $20 \times$ microscope objective (MO) and a lens with a focal length of $f=125$ mm, which is then modulated in amplitude by a Holoeye LC2012 transmission spatial light modulator (SLM). This enables the generation of an arbitrary amplitude pattern projected onto the beam, facilitating the simultaneous excitation of multiple waveguides. During this part, the intensity and size of the light spot (disk) can be configured for each waveguide under study. Once the image amplitude profile is formed, the phase structure is added by means of a Holoeye PLUTO reflection SLM. An example of an input image condition, formed by amplitude and phase modulation, is shown in Fig.~\ref{setupimg}(b1). Upon completing the modulation process, the image is then optically transferred to the input facet of the photonic chip by using a telescope and a $4 \times$ MO. An example of an output profile is shown in Fig.~\ref{setupimg}(b2), after the input condition has traveled along a 70 mm graphene photonic lattice.


\vskip 0.8cm
\noindent{\bf {5. Experimental analysis of trivial and topological graphene ribbon lattices.}}
\vskip 0.5cm

\begin{figure}[htbp]
\centering\includegraphics[width=0.99\textwidth]{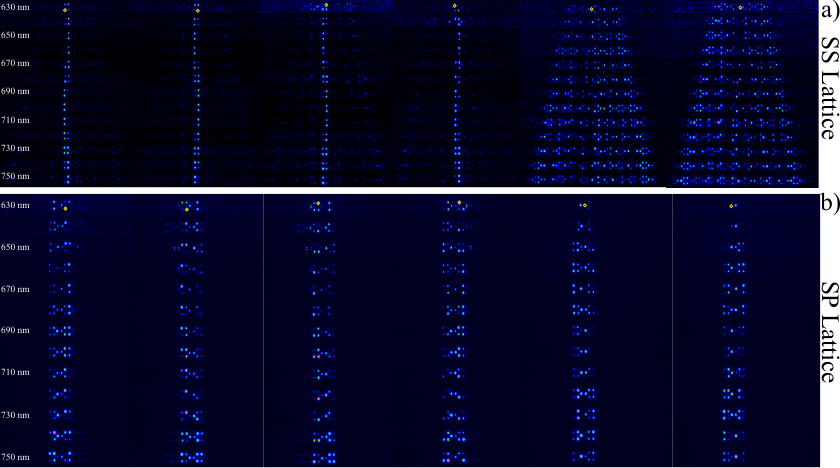}
\caption{Output profiles, at different wavelengths, after single site bulk excitations of the unit cell for (a) SS and (b) SP lattices. Yellow circles indicate the input site. From left to right: B, C, E, F, A and D input sites, respectively.}
\label{bcefssbulk}
\end{figure}

{\it Single Site Bulk excitation}. We first explore the strong contrast in between the dynamical properties of the trivial (SS) and topological (SP) photonic lattices. We start by exciting the unit cell at the lattice bulk using different input wavelengths coming from a SC laser source. Fig.~\ref{bcefssbulk} shows a compilation of several output profiles after exciting sites A, B, C, D, E, and F of a (a) trivial SS and (b) nontrivial SP lattices. For the SS case, we observe flat band oscillation and weak transport after exciting sites B, C, E, and F, and ballistic transport only after exciting sites A and D. On the contrary, for the SP case, the light is mostly confined at two neighbouring unit cells due to the excitation of a perfectly flat band spectrum only, independent of the input site. 


\begin{figure}[htbp]
\centering\includegraphics[width=.9\textwidth]{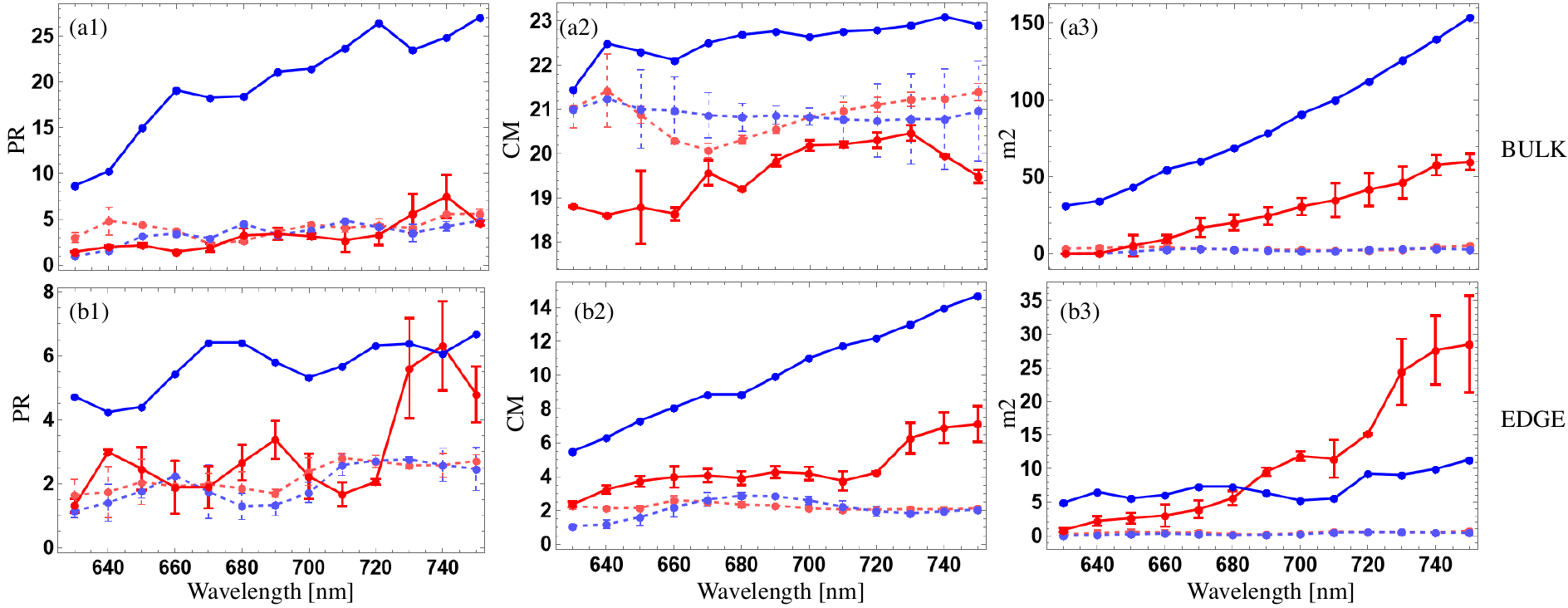}
\caption{Participation ratio (PR), Center of Mass (CM) and second moment (m2) for SS (full lines) and SP (dashed lines) lattices, and for (a) bulk and (b) edge excitation. A-site (B and F) excitation is shown in blue (red) color.}
\label{quantify}
\end{figure}

We analyze the obtained data by computing the Participation Ratio (PR), center of mass (CM), and second moment ($m_2$), for each image shown in Fig.~\ref{bcefssbulk}, and collect results in Fig.~\ref{quantify}(a). In the SS lattice case, after exciting central A and D sites (full-blue curves) we observe that the PR and $m_2$ increase upon wavelength, as an indication of the ballistic transport originated from a dispersive-only band excitation. On the other hand, after exciting sites B,C,E or F (see full-red curves) we notice a slight oscillation in the PR, and a slower increment of $m_2$. This indicates a mixture in between the excitation of dispersive and flat bands. The CM in Fig.~\ref{quantify}(a2) does not exhibit a significant increment upon wavelength, as the energy spreads out symmetrically from the lattice center. On the other hand, the excitation of a SP nontrivial lattice [see dashed blue and dashed red data in Fig.~\ref{bcefssbulk}(a)] produces an oscillatory bounded behavior, where the PR, CM and $m_2$ oscillate for an increasing wavelength. This is in agreement with the caging phenomena shown in Fig.~\ref{bcefssbulk}(b), and directly confirms the flat-only band spectrum for a nontrivial flux. 


\newpage

\begin{figure}[htbp]
\centering\includegraphics[width=0.9\textwidth]{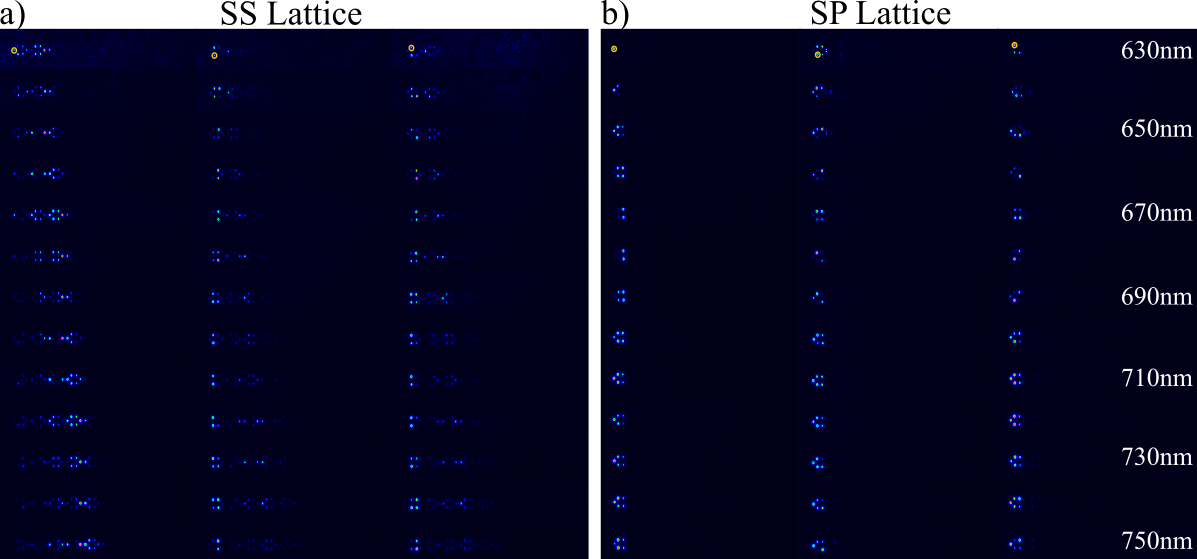}
\caption{Output profiles after excitation of sites A, B and F, for the first (left-edge) unit cell, for different input wavelengths, and for (a) SS and (b) SP lattices. Yellow circle indicates the input site.}
\label{bcefssedge}
\end{figure}

{\it Single Site Edge excitation}. Now, we study the excitation of the lattice edges at different wavelengths. For simplicity, we compare the excitation of sites A, B and F at the first unit cell (left-edge) for both, SS and SP, cases. Fig.~\ref{bcefssedge} compiles our results, with the output profiles for different excitation wavelengths. For a SS lattice, we observe that an A site excitation produces ballistic transport as a result of exciting the dispersive part of the spectrum only. Data analysis in Fig.~\ref{quantify}(b) (see full-blue curve) shows a clear increment of the CM, with a weakly increment of the PR and a rather constant $m_2$. However, if we compare this result with the excitation of sites B and F, we clearly notice the competition in between the flat (tendency to trap the energy at the excited unit cell) and the dispersive (discrete diffraction) bands. The analysis in Fig.~\ref{quantify}(b) (see full-red curve) shows an oscillating PR, a slowly increment of CM, and a rather fast increment on $m_2$. The edge excitation of an SP ribbon lattice, having a nontrivial flux of $\pi$, shows remarkable differences compared to a SS lattice. We clearly observe in Fig.~\ref{bcefssedge}(b), for the three excitation sites, that the light remains trapped/caged and oscillating at the first unit cell only, as a result of the direct excitation of the three edge states existing at this regime. We notice a quite low background for all wavelengths, due to the fact that the light keeps strongly localized at the edge unit cell. The analysis shown in Fig.~\ref{quantify}(b) (see blue and red dashed lines) indicates a small and weakly oscillating PR, a rather constant CM, and a very small $m_2$, indicating absence of transport. These results show the robustness of the edge localization for a wide range of wavelengths. 

\vskip 1cm
\noindent{\bf {6. Excitation of zero energy edge states.}}
\vskip 0.5cm

\begin{figure}[htbp]
\centering\includegraphics[width=1.\textwidth]{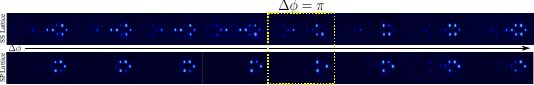}
\caption{Output profile for the simultaneous excitation of A,C and E sites with different phases in between the A and C/E sites. The phase is changed at site A only, while keeping constant the phase at C/E sites. The horizontal arrow indicates the increment of the phase difference $\Delta\phi$.}
\label{edgephase}
\end{figure}

{\it Robustness to the change of input phase}. We demonstrate the robustness of the edge state in $\pi$-graphene lattices by applying a three-site input condition using the image setup at $640$ nm. In Fig.~\ref{edgephase} we observe the flux $\Phi=\pi$ effect for the simultaneous excitation of A,C and E sites. Similar to the case of a single-site excitation shown in Fig.~\ref{bcefssedge}, the light remained trapped inside the unit cell for the SP lattice. A precise excitation of the topological edge state is achieved around a phase difference $\Delta\phi=\pi$ in between A and C/E sites. This behavior demonstrates the robustness of the edge state to a phase change in an SP topological lattice. In contrast, the excited pattern in the SS lattice shows dispersive-like behavior, as shown for $\Delta\phi=\pi$ in Fig.~\ref{edgephase}. This clearly show the absence of the edge modes and edge localization for $0$-flux ribbon graphene lattices.

\begin{figure}[htbp]
\centering\includegraphics[width=1.0\textwidth]{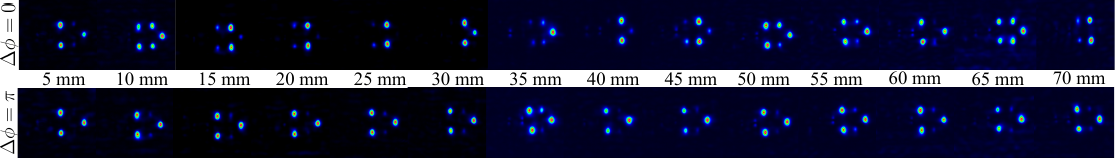}
\caption{Output profile versus propagation coordinate $z$ for a simultaneous excitation of sites A, C and E, with an in-phase structure $\Delta\phi=0$ (top) or an out of phase structure $\Delta\phi=\pi$ (bottom).}
\label{propagation}
\end{figure}

{\it Steady edge mode propagation}. Finally, we study the dynamical stability for the edge compact topological mode along the propagation coordinate $z$ for a topologically nontrivial SP lattice. We first excite a three (A, C, and E) sites input pattern having a $\Delta\phi=0$ input phase and show compiled results in Fig.~\ref{propagation}-top. For this in-phase initial condition the light does not remain trapped at the excitation sites and oscillates at the first unit cell, similar to the single-site excitation shown in Fig.\ref{bcefssedge}(b). An out of phase input excitation, corresponding to a perfectly compact edge topological state profile, shows how this mode remains perfectly compact along the propagation coordinate, with only slight fluctuations associated with fabrication errors (every image is obtained for a different sample~\cite{spirodrigo}). This compiled results demonstrate the robustness and stability of the edge topological state along propagation, as different samples introduce a different amount of disorder.

